\documentclass[aps,reprint,prd,nofootinbib]{revtex4-1}

\usepackage{bm}
\usepackage{graphicx} %
\usepackage{amsmath} %
\usepackage{amssymb} %
\usepackage{url}
\usepackage{subfigure} %
\usepackage{float} %
\usepackage{upgreek} %
\usepackage{multirow} %
\usepackage{color} %
\usepackage{lineno} %
\usepackage{hyperref} %
\usepackage{booktabs}
\usepackage{placeins}
\usepackage{ulem}
\usepackage[usenames,dvipsnames]{xcolor}
\hypersetup{colorlinks=true, urlcolor=black, linkcolor=blue, citecolor=red} %

\usepackage{amsmath}
\usepackage{tcolorbox}
\allowdisplaybreaks[4]

\begin{document}

\title{Covariant transverse-traceless projection for secondary gravitational waves}

\author{Atsuhisa Ota${}^{1,2}$}
\email{iasota@ust.hk}

\author{Hayley J. Macpherson${}^3$}
\email{h.macpherson@damtp.cam.ac.uk}

\author{William R. Coulton${}^4$}
\email{wcoulton@flatironinstitute.org}

\affiliation {${}^1$HKUST Jockey club Institute for Advanced Study, The Hong Kong University of Science and Technology, Clearwater Bay, Hong Kong, P.R.China } 
\affiliation {${}^2$Department of Physics and Astronomy Ohio University, Athens, OH, 45701, United States}
\affiliation{${}^3$Department of Applied Mathematics and Theoretical Physics, Cambridge CB3 0WA, UK}
\affiliation{${}^4$Center for Computational Astrophysics, Flatiron Institute, New York, NY 10010, United States}

\date{\today}

\begin{abstract}

Second-order tensor modes induced by nonlinear gravity are a key component of the cosmological background of gravitational waves. A detection of this background would allow us to probe the primordial power spectrum at otherwise inaccessible scales. 
Usually, the energy density of these gravitational waves is studied within perturbation theory in a particular gauge --- a connection between our physical spacetime and a fictitious background. It is a widely recognized issue that the second-order, scalar-induced gravitational waves are gauge dependent.
This issue arises because they are not well-defined as tensors in the physical spacetime at second-order and are thus unphysical.
In this paper, we propose the covariant transverse-traceless projection of the extrinsic curvature to study cosmological gravitational waves on a spatial hypersurface.
We define a new energy density which is based purely on spacetime tensors, independent of perturbation theory, and thus is gauge invariant by definition. 
We show that, in the context of second-order perturbation theory, this new energy density contains only propagating modes in the constant-time hypersurface in the Newtonian gauge. We further show that we can recover the same gravitational waves after a transformation to the synchronous gauge, so long as we correctly identify the Newtonian hypersurface.


\end{abstract}

\maketitle


\section{Introduction}
A new era of observational astrophysics began with the discovery of gravitational waves (GWs) from a binary black hole merger \citep{LIGOScientific:2016aoc}. Increasing numbers of events will eventually allow for precise measurements of cosmological parameters using GWs~\citep{Abbott2021}. In addition to measurements of isolated events, both astrophysical and cosmological backgrounds of GWs may be detected in the coming years~\cite{Lentati:2015qwp, NANOGrav:2015aud,LIGOScientific:2016wof, LIGOScientific:2017zlf, Audley:2017drz, NANOGrav:2020bcs}.
A cosmological background of GWs is usually described by tensor perturbations on top of a background Friedmann-Lema\^itre-Robertson-Walker (FLRW) spacetime.
While the primordial GWs generated during inflation is the best known example of such a 
GW background~\cite[see, e.g.][]{Dodelson:2003ft,Mukhanov:2005sc,Weinberg:2008zzc},
nonlinear dynamics of gravity also source propagating tensor modes at second-order in perturbation theory
~\cite{Matarrese:1993zf,
Matarrese:1997ay,
Mollerach:2003nq,
Ananda:2006af,
Baumann:2007zm,
Gong:2019mui,
Domenech:2019quo,
Zhou:2021vcw}.
These scalar-induced GWs are attracting growing attention as a means to probe large density perturbations in the early Universe~\cite{Saito:2008jc,
Saito:2009jt,
Assadullahi:2009nf,
Alabidi:2013wtp,
Nakama:2016gzw,
Kohri:2018awv,
Unal:2018yaa,
Inomata:2018epa,
Cai:2018dig,
Cai:2019amo,
Inomata:2019zqy,
Inomata:2019ivs,
Cai:2019elf,
Cai:2019cdl,
Ota:2020vfn,
Inomata:2020lmk,
Pi:2020otn,
Domenech:2020kqm,
Dalianis:2020cla,
Yuan:2020iwf,
coulton:2020,
Atal:2021jyo,
Adshead:2021hnm,
Domenech:2021ztg}.

Cosmological perturbation theory has a gauge freedom, which arises from general covariance of the physical spacetime.
Induced tensor modes have been studied in a variety of gauges, and have been shown to contain gauge-dependent, non-propagating tensor modes at second order~\cite{Hwang:2017oxa,Yuan:2019fwv,Tomikawa:2019tvi,DeLuca:2019ufz,Inomata:2019yww,Ali:2020sfw,Chang:2020mky,Domenech:2020xin,Chang:2020iji,Cai:2021ndu,Cai:2021jbi}. 
This implies the effective GW energy density usually defined in the literature is gauge dependent. This is a crucial issue for writing the GW effective energy momentum as a spacetime tensor in the physical spacetime.
We therefore must find a definition of the induced GW energy density that is consistent at nonlinear order.

From a perturbation theory perspective, the construction of gauge-invariant quantities at nonlinear order has been discussed~\cite{Malik:2003mv,Nakamura:2003wk,DeLuca:2019ufz}.
A gauge-invariant version of the nonlinear Isaacson formalism may be possible using such gauge-invariant variables~\cite{Abramo:1997hu}.
However, this approach is complex at nonlinear order, and the existence of the GW energy momentum as a spacetime tensor is not guaranteed.
Refs.~\cite{Cai:2021ndu,Cai:2021jbi} recently introduced a reference manifold to define the GW energy density as a quasi-local energy density --- which may offer alleviation of the gauge issue ---; however, here, we consider an alternate approach.
In this paper, we propose a new way to define the GW energy density in a covariant perspective and without reference to any background spacetime.
Such an approach is, by definition, free from the gauge issue --- making it potentially viable for non-perturbative analysis, e.g., using numerical relativity (NR) to study nonlinear regimes such as primordial black holes or cosmological structure formation.

Latin indices $(a,b,c\cdots)$ represent abstract indices for a general tensor field, Greek indices ($\mu,\nu,\rho,\cdots$) are spacetime indices with a coordinate basis and take values $0\ldots3$, and Latin indices ($i,j,k,\cdots$) are spatial indices and take values $1\ldots3$. Repeated indices imply summation irrespective of their position, and indices are always raised and lowered by the spatial or spacetime metric tensor. In Appendix~\ref{index:spp} we provide further specifics on our index convention.
We set the speed of light $c=1$ throughout this paper.

\section{Gravitational wave energy in the covariant perspective}

GWs are usually defined in cosmological perturbation theory as the transverse-traceless~(TT) part of the spatial metric for a particular background spacetime. 
This procedure is not covariant with respect to a general coordinate transformation, which causes the GWs to be gauge dependent at second-order in cosmological perturbation theory.
From a geometrical perspective, using a fully covariant decomposition in the physical spacetime should be more robust.
In Ref.~\cite{York:1974psa}, York discussed such a covariant TT decomposition in 3--space, which we will apply to study the GW energy density in cosmology.

\subsection{York's covariant TT projection}\label{subsec:york}

Ref.~\cite{York:1974psa} showed that the \textit{covariant} TT decomposition for an arbitrary symmetric tensor $Q^{ab}$, on a smooth Riemannian 3--manifold with metric $\gamma_{ab}$, is written as
\begin{align}
	Q^{ab} = Q_{\rm TT}^{ab} + (LV)^{ab} + \frac{1}{3}\gamma^{ab}Q,\label{decom}
\end{align}
where $Q\equiv \gamma_{ab}Q^{ab}$ is the trace of $Q^{ab}$, and we have defined
\begin{align}
	(LV)^{ab} \equiv D^a V^b + D^b V^a - \frac{2}{3}\gamma^{ab}D_cV^c,\label{LL}
\end{align}
where $D_a$ is the covariant derivative associated with $\gamma_{ab}$.
$V^a$ is the unique solution to the vector Laplacian equation, namely,
\begin{align}
	D_a (LV)^{ab} =  D_a \left( Q^{ab} - \frac{1}{3}\gamma^{ab}Q\right),\label{veceq} 
\end{align}
for certain boundary conditions~\cite{York:1974psa}.
The covariant TT conditions on $Q_{\rm TT}^{ab}$ imply $D_b Q_{\rm TT}^{ab} = 0$ and  $\gamma_{ab} Q_{\rm TT}^{ab}=0$, which we note are distinct from the \textit{non-covariant} TT conditions usually considered for the tensor perturbation in an FLRW background spacetime.

Since the TT condition \eqref{decom} defines a tensor which is TT with respect to $\gamma_{ab}$, we cannot apply this covariant decomposition to the spatial metric itself.
Therefore, we must first identify a tensor that carries the energy of GWs in a covariant perspective.
In NR simulations of binary black holes, the covariant TT part of the extrinsic curvature is loosely associated with GWs. Specifically, this tensor is often set to zero in the initial data to remove any GWs not generated by the binary itself~\citep[see Chapter 3 of][]{BaumgarteShapiro2010}. We will consider this tensor in the following section and assess its use in cosmology.

York's construction implies that we first need to define a 3--dimensional spatial foliation of a 4--dimensional spacetime; it requires choosing a particular 
spatial hypersurface.
The hypersurface dependence of GWs defined as spin--2 degrees of freedom (DOFs) associated with the spatial metric is, therefore, essentially inevitable. We discuss the implications of this hypersurface dependence in Section~\ref{cttime}. 
We note that `hypersurface dependence' is distinct from `gauge dependence': the latter is defined only in the context of perturbation theory and the former refers to the dependence on a particular foliation of spacetime into a series of spatial surfaces, which is independent of perturbation theory~\cite{Nakamura:2020pre}.

\subsection{Curvature energy density in FLRW spacetime}

In Section~\ref{sec:HGW} below, we will define the energy density of GWs from their contribution to the total energy density via the Hamiltonian constraint. In this section, we will first briefly review the analogous (but familiar) contribution of the curvature to the total energy density in an FLRW spacetime. 

In the FLRW model, the metric tensor in spherical coordinates is
\begin{align}
	ds^2 = -dt^2 + a^2 \left[\frac{dr^2}{1-\kappa r^2} + r^2 (d\theta^2+\sin^2\theta d\phi^2)\right],\label{flrw:metric}
\end{align}
where $a=a(t)$ is the scale factor, and $\kappa$ is the scalar curvature. The total (critical) energy density, $\bar \rho_{\rm c}$, is defined from the Hamiltonian constraint, which for an FLRW spacetime reduces to the Friedmann equation 
\begin{align}
3M_{\rm pl}^2H^2 = \bar \rho_{\rm c},	
\end{align}
where $H$ is the Hubble parameter, $M_{\rm pl}=1/\sqrt{8\pi G}$ is the reduced Planck mass, and $G$ is the gravitational constant.
In the above, $\bar \rho_{\rm c}$ is distinct from the energy density of the matter content, $\bar\rho$.
The remaining energy density is the contribution from curvature
\begin{align}
    \bar \rho_{\rm c} - \bar \rho = \rho_\kappa = -\frac{\kappa}{ a^2} .\label{energy:curve}
\end{align}
Here we see that the curvature in the metric tensor \eqref{flrw:metric} contributes to the total energy density and therefore to the cosmic expansion, $H$.
We do not need to consider a perturbative expansion with respect to $\kappa$ to define the energy-density of curvature in Eq.~\eqref{energy:curve}.

In a perturbed FLRW spacetime, GWs are also present in the components of the metric tensor, and will therefore contribute to the cosmic expansion. This should also hold in the case of a general spacetime --- independent of a background cosmology --- which will also contain GWs.
In the next section, we will show that the energy density in the Hamiltonian constraint for a general spacetime naturally includes both curvature and GW energy densities.

\subsection{Hamiltonian constraint for GW energy density}\label{sec:HGW}

We consider a 4--dimensional spacetime with metric tensor $g_{ab}$, which we foliate into a series of 3--dimensional spatial hypersurfaces with time-like unit normal $n_a$.  
The extrinsic curvature of the hypersurfaces, $K_{ab}$, is the Lie derivative of the spatial metric
\begin{align}
    \gamma_{ab} \equiv g_{ab} + n_an_b,	\label{defgamma6}
\end{align}
along the normal vector, namely
\begin{align}
	K_{ab} \equiv \frac12\pounds_n \gamma_{ab}\label{defKab11}.
\end{align}
The trace of the extrinsic curvature $K\equiv \gamma^{ab}K_{ab}$ is the logarithmic expansion rate of the volume element $\sqrt{|\gamma|}$ 
along $n^a$, which reduces to $K=3H$ in an FLRW spacetime.
The Friedmann equation comes from the Hamiltonian constraint~\cite[see, e.g.][]{Gourgoulhon:2007ue}
\begin{equation}
    R + K^2 - K_{ab}K^{ab} - \frac{2\rho}{M_{\rm pl}^2} = 0,\label{eq:Hconstraint}
\end{equation}
where $R$ is the 3--Ricci curvature of the hypersurface, and $\rho\equiv T_{ab}n^a n^b$ is the energy-density of matter, with $T_{ab}$ the energy-momentum tensor.
We can recast Eq.~\eqref{eq:Hconstraint} into a Friedmann-like form, namely
\begin{align}\label{eq:ham_recast}
	 3M_{\rm pl}^2\left(\frac{K}{3}\right)^2 = \rho_{\rm c}, 
\end{align}
where $\rho_{\rm c}$ is the total energy density and $\rho_{\rm c} - \rho $ therefore must contain all forms of energy density not contained in matter, including 
the energy density of GWs. 
As we mentioned above, we will show that the covariant TT part of the extrinsic curvature may represent the kinetic energy density of the GWs in a covariant perspective. 
We aim to isolate the GWs from other contributions by considering the covariant TT decomposition \eqref{decom} of the extrinsic curvature~(see Ref.~\cite{Clough2018} for a similar method without the transverse projection), namely
\begin{align}
	K^{ab} = K_{\rm TT}^{ab} + (LV)^{ab} + \frac{1}{3} \gamma^{ab}K. \label{covAab}
\end{align}
Combining this decomposition with Eq.~\eqref{LL}, we find
\begin{align}
    \label{eq:KK}
\begin{split}
    K_{ab}K^{ab} &= K^{\rm TT}_{ab}K^{ab}_{\rm TT} + \frac{1}{3}K^2  \\
    &+ (LV)_{ab}(LV)^{ab} + 4D_a (V_b K^{ab}_{\rm TT}),
\end{split}
\end{align}
and substituting Eq.~\eqref{eq:KK} into Eq.~\eqref{eq:Hconstraint}, we can now write 
Eq.~\eqref{eq:ham_recast} as 
\begin{align}
    3 M_{\rm pl}^2 \left (\frac{K}{3}\right)^2 =& \rho + \rho_{K} + \rho_{R}+ \rho_{V}+\rho_{\rm div}, \label{eq:ham_recast_full}
\end{align}
where we have defined
\begin{align}
	\rho_{K} &\equiv \frac{M_{\rm pl}^2}{2} K^{\rm TT}_{ab} K^{ab}_{\rm TT}, \label{eq:rhoK} \\
	\rho_{R} &\equiv -\frac{M_{\rm pl}^2}{2} R, \label{eq:rhoR} \\
	\rho_{V} &\equiv \frac{M_{\rm pl}^2}{2}(LV)_{ab}(LV)^{ab}, \label{eq:rhoL} \\
	\rho_{\rm div} &\equiv 2M_{\rm pl}^2 D_a (V_b K^{ab}_{\rm TT}). \label{eq:rhodiv} 	
\end{align}
The energy density in Eq.~\eqref{eq:rhodiv} is a covariant divergence, which 
vanishes in the perturbative approach when taking the Brill-Hartle average, as shown in Ref.~\cite{Isaacson:1968zza} \citep[see also Section~35.15 of][]{MTW}. In the general case, this becomes a surface term and will thus still vanish when considering its spatial average.
The left hand side of Eq.~\eqref{eq:ham_recast_full} is the energy of expansion of the hypersurface, and $\rho_{K}+\rho_{R}+\rho_{V}$ is the contribution to this expansion from the metric tensor.

The energy density $\rho_{R}$ must contain curvature as well as the gradient energy density of GWs --- since $R$ contains only \textit{spatial} gradients of the metric tensor --- and $\rho_{V}$ is related to the vector field $V^a$, including the scalar shear, and thus its physical interpretation is unclear. However, since the latter is not purely TT it cannot contain GWs energy in a covariant perspective. 

The energy density $\rho_K$ \textit{is} purely TT and is built from time derivatives of the metric tensor, and we thus define $\rho_{K}$ as the kinetic energy density of GWs\footnote{This definition holds after performing a proper volume average of $\rho_K$ over a domain larger than the wavelength of the GWs of interest}.
We have confirmed that Eq.~\eqref{eq:rhoK} reduces to the Isaacson energy density~\cite{Isaacson:1968hbi,Isaacson:1968zza,Maggiore:1999vm} in the case of linear tensor perturbations about an FLRW background cosmology,~\footnote{To be more precise, $\rho_K=\rho_{\rm Isaacson}/2$, and $\rho_K$ corresponds to the time derivative term when we do not use the linearized equation of motion for the tensor mode in Isaacson's derivation. When $\langle \rho_K\rangle = \langle \rho_R\rangle$, our expression recovers $\langle \rho_{\rm Isaacson}\rangle = \langle \rho_K\rangle + \langle \rho_R\rangle $.}
and we also find $\rho_K=0$ for linear scalar or vector perturbations, which can be seen from our perturbation theory calculations in Section~\ref{sec:2nd-order}.
Therefore, we might naively expect the 4--scalar $\rho_{K}$ to be a generalized kinetic energy density of GWs which could be useful in cosmology.
Since $\rho_{K}$ is a scalar in the spacetime, any coordinate transformation cannot set $\rho_{K}=0$, unlike the tensor mode in perturbation theory.

For linear tensor modes, $\rho_R$ reduces to the gradient energy of GWs with time average the same as that of $\rho_K$ at high frequencies. 
In the case of linear scalar perturbations, $\rho_R$ contains only the curvature energy density. 
Therefore, in a general spacetime $\rho_R$ will contain \textit{both} curvature and GW gradient energy density, making it difficult to isolate the GW contribution.
Instead using $\rho_K$ to characterize the GW energy density, we can potentially separate the GW energy from other sources.
Further, the notion of gauge-dependence only arises within perturbation theory.
Since Eq.~\eqref{eq:ham_recast_full} is exact and obtained in the physical spacetime without reference to perturbation theory, $\rho_{K}$ must be gauge independent by definition.

\section{Gauge transformation}
In practice, we need perturbation theory for the analytic evaluation of $\rho_{K}$, and hence the gauge issue arises. 
In Section~\ref{sec:gauge_freedom} we explain the gauge freedom in cosmological perturbation theory and in Section~\ref{sec:GWgauge} we provide a proper interpretation of the gauge transformation of GWs.

\subsection{Gauge freedom in perturbation theory}\label{sec:gauge_freedom}
\begin{figure}
	  \includegraphics[width=\linewidth]{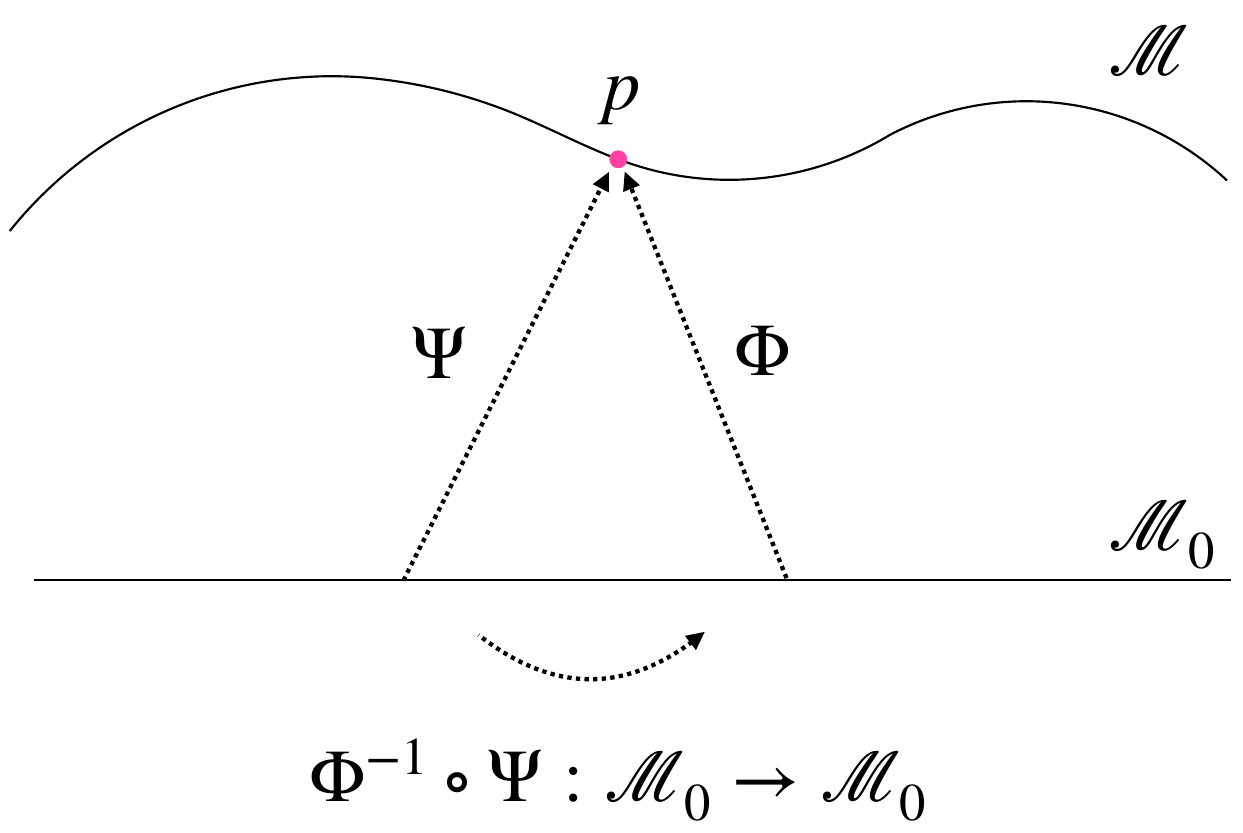}
  \caption{Illustration of gauge transformation on a background spacetime $\mathcal M_0$. There exits an arbitrary pair of identifications $\Phi$ and $\Psi$ between the physical spacetime $\mathcal M$ and a background manifold $\mathcal M_0$ due to general covariance on the physical spacetime. Different points $\Phi^{-1}(p)$ and $\Psi^{-1}(p)$ on $\mathcal M_0$ represent the same point $p$ on $\mathcal M$, so that the diffeomorphism $\Phi^{-1}\circ\Psi:\mathcal M_0 \to \mathcal M_0$ is a non physical freedom, that is, the gauge freedom.
  } 
  	\label{fig}
\end{figure}

In cosmological perturbation theory, we start from a physical spacetime $\mathcal M$ with some approximate symmetry --- e.g. homogeneity and isotropy on large scales.
Then we consider a solution to Einstein's equations which has that exact symmetry --- e.g., the FLRW model --- and define this as the background spacetime $\mathcal M_0$. 
We then identify this fictitious background spacetime with the physical spacetime through a particular choice of \textit{gauge}, $\Psi$: $\mathcal M_0\to \mathcal M$~(see Fig.~\ref{fig}).
This identification is not unique, which is the ``gauge freedom'' in perturbation theory~(see, e.g., Ref.~\cite{Nakamura:2020pre} for a recent review).

\medskip
A physical quantity is expressed by a spacetime tensor $Q$ in $\mathcal M$. When the calculation of $Q$ is difficult in practice, we often go to $\mathcal M_0$, where $Q$ is identified with $\Psi^*Q$; the pullback of $Q$ by $\Psi$.
However, a choice of gauge is not unique, and one may choose $\Phi: \mathcal M_0\to \mathcal M$ and $\Phi^*Q$ instead.
Both $\Psi^*Q$ and $\Phi^*Q$ are physically equivalent, but their representations are not necessary the same.
The diffeomorphism $\Phi \circ \Psi^{-1}: \mathcal M_0\to \mathcal M_0$ is the gauge transformation, and the representation of $Q$ changes from $Q_0\equiv \Psi^*Q$ to $\hat Q_0\equiv \Phi^*Q$.
In general, the gauge transformation is approximated by the Knight-diffeomorphism: a sequence of exponential maps generated by a set of infinitesimal tangents in the 4--dimensional spacetime. 
The gauge transformation of $Q_0$ along $\xi^{a}_{(1)}, \xi^{a}_{(2)}, \cdots$ is written as~\cite{Sonego:1997np}
\begin{align}
	Q_0 \to \hat Q_0 = e^{\pounds_{\xi_{(1)}}}e^{\pounds_{\xi_{(2)}}}\cdots Q_0.\label{gaugetf}
\end{align}
For the remainder of this paper, we use a hat to denote quantities transformed by Eq.~\eqref{gaugetf}.
The tensor $Q_0$ is identified with $\hat Q_0$ by the Knight diffeomorphism, which implies that they are physically equivalent quantities.
Therefore, the variation
\begin{align}
	\delta Q_0\equiv \hat Q_0 - Q_0,
\end{align}
represents unphysical degrees of freedom, that is, the gauge freedom.

\subsection{Gauge transformation of GWs}\label{sec:GWgauge}

If there exists a generalised energy density of GWs, it could be a 4--scalar --- associated with either a hypersurface or an observer --- or the time-time component of a 4--tensor. 
In either case, such a quantity is expressed via a tensor $Q$ in $\mathcal M$.
In a background spacetime, we can caclualte $\Psi^*Q$ or $\Phi^*Q$, and we may consider the gauge transformation of those tensors by using Eq.~\eqref{gaugetf} directly, which guarantees their gauge-invariance at the leading order~(see the proposition 1 of Ref.~\cite{Bruni:1996im}).
Therefore, the leading order gauge invariance is a necessary requirement when constructing the effective energy density of GWs in a background spacetime, $\Psi^*Q$, without having $Q$.
The gauge dependence of the second-order tensor modes implies that we cannot simply apply Isaacson's formula to guarantee the leading order gauge invariance of the energy density of scalar-induced GWs.
One method to overcome this issue is to introduce a second-order, gauge-invariant tensor perturbation and thus maintain Isaacson's formula, as considered by Ref.~\cite{DeLuca:2019ufz}.
A second approach would be to find a generalized energy density such that, within perturbation theory, the gauge dependence of the tensor modes cancels to realize the gauge invariance at leading order.
In this paper, we make use of this second approach.

\medskip 
Our solution is simple: as discussed in the previous section, we use $\rho_K$ --- which is a 4--scalar --- to define the energy density of induced GWs. 
Eq.~\eqref{gaugetf} straightforwardly applies to $\Psi^*\rho_K$ in a background spacetime, which we simply denote by $\rho_K$ when in the context of perturbation theory.
Hereafter, other tensors in $\mathcal M$ are always understood as tensors in $\mathcal M_0$ in a similar way. 
The gauge transformation of $\rho_{K}$ is
\begin{align}
	\hat \rho_{K} = \rho_{K} + \xi^{a}_{(1)}\nabla_a \rho_{K}+\cdots,\label{rhoTTgt}
\end{align}
where $\cdots$ represents corrections higher order in $\xi^a_{(1)}, \xi^a_{(2)}\cdots$. 
For the scalar-induced GWs, the leading-order term of $\rho_{K}$ is fourth order in the scalar perturbations, so we find that 
\begin{align}
\label{igwgaugeinv}
    \hat \rho_{K} = \rho_{K} 	
\end{align}
is satisfied in any gauge to fourth order in scalar perturbations, which is the main result of this paper.
We have proposed an energy density of GWs that is expressed by a 4--scalar which is gauge invariant at leading order.
We will provide the explicit form of $\rho_K$ in perturbation theory in Section~\ref{sec:2nd-order} and will show that the gauge-dependent part 
cancels in $\hat \rho_K$.

\begin{figure}
	  \includegraphics[width=\linewidth]{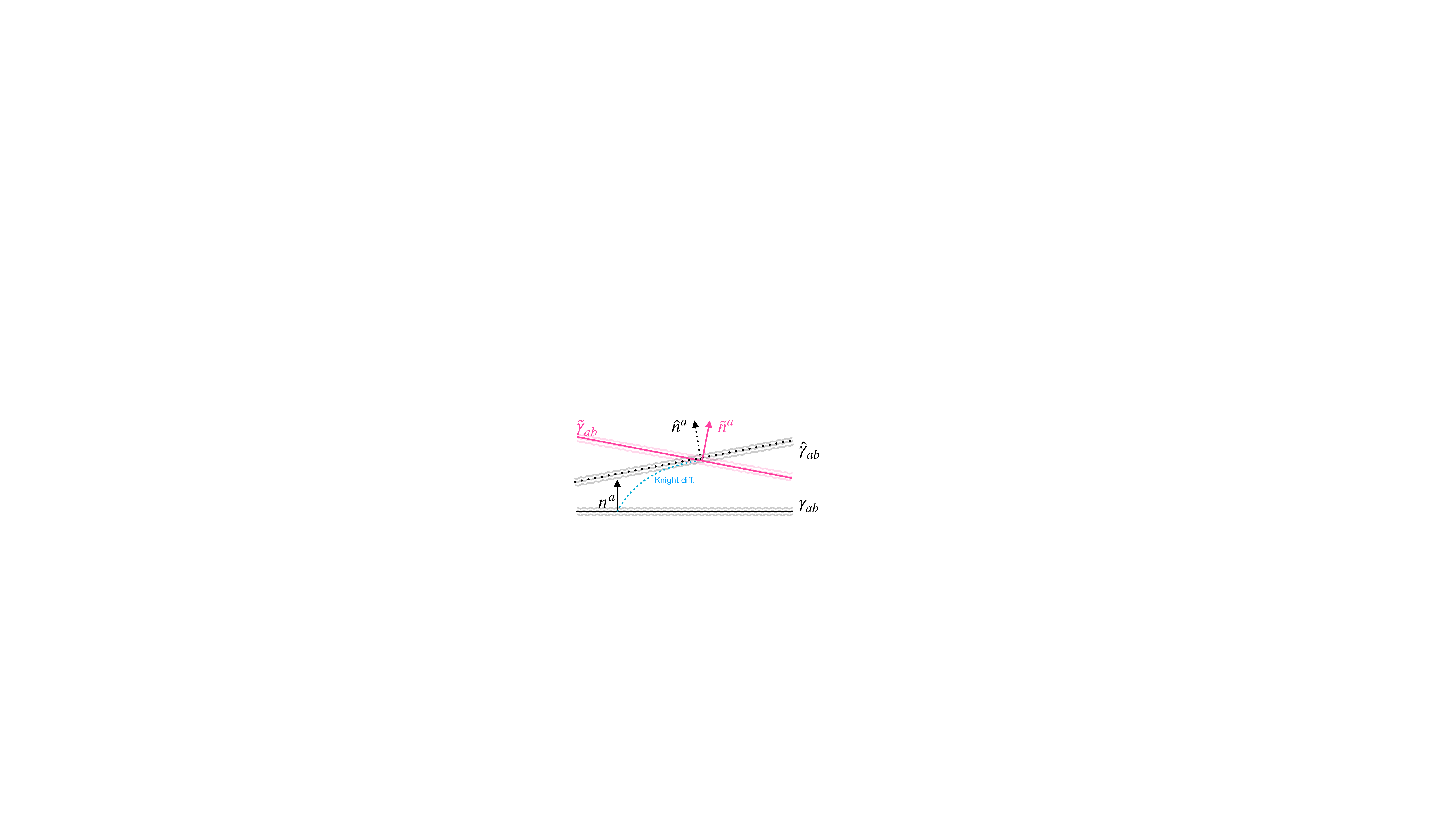}
  \caption{Here we illustrate perturbatively defined spatial metrics $\gamma_{ab}$, $\tilde \gamma_{ab}$, and $\hat \gamma_{ab}$  in a background manifold. 
  Points connected by the Knight diffeomorphism (blue dotted curve) represent the same point in the physical manifold. 
  A constant-time hypersurface in the new gauge,  $\tilde \gamma_{ab}$, is different from the transformed hypersurface $\hat \gamma_{ab}$ with normal $\hat n^a$. Physically equivalent GWs are shown in gray, which are distinct from those shown in magenta.} 
  	\label{fig1}
\end{figure}

\section{Hypersurface dependence}\label{cttime}

Eq.~\eqref{igwgaugeinv} implies that we can compute $\rho_K$ for a specific hypersurface in any gauge.
The gauge transformation of $\rho_K$ is given by Eq.~\eqref{rhoTTgt}, which identifies $\hat \rho_{K}$ with the energy density defined in the original hypersurface in the new gauge.
A practical choice of hypersurface might be one where the time coordinate is constant.
A gauge transformation will change the constant-time hypersurface and therefore the corresponding energy densities are physically different scalars associated with different spatial hypersurfaces.
Here we denote the former as $\rho_K$ and the latter as $\tilde \rho_K$, whereas the gauge transform of $\rho_K$ is $\hat \rho_K$. 

As we already discussed, $\rho_K$ and $\hat \rho_K$ are physically equivalent and their representations are equal at leading order. 
However, as we will show in the following, in general we have $\hat \rho_K \neq \tilde \rho_K$. This is because gauge dependence and hypersurface dependence are different concepts. 
The former can be imposed from theoretical consistency, but the latter depends on our choice of the system.

\medskip
To show that $\hat \rho_K \neq \tilde \rho_K$ in general, we start with
$\hat \rho_K$ which is the energy density associated with the time-like normal in the new gauge 
\begin{align}
	\hat n_a =e^{\pounds_{\xi_{(1)}}}e^{\pounds_{\xi_{(2)}}}\cdots n_a.
\end{align}
The construction of the extrinsic curvature is given coordinate independently, so once we have $\hat n_a$ we straightforwardly obtain
\begin{align}
    \hat{K}_{ab}  =\frac{1}{2} \pounds_{\hat n}  \hat \gamma_{ab},\label{kab22}
\end{align}
where the spatial metric in the new gauge is
\begin{align}
    \hat \gamma_{ab}= \hat g_{ab}+\hat n_a \hat n_b.
\end{align}
Then, $\hat \rho_K$ can be constructed from $\hat K^{\rm TT}_{ab}$ (which is TT with respect to $\hat \gamma_{ab}$).
We also arrive at Eq.~\eqref{kab22} when directly considering the gauge transformation of $K_{ab}$ via Eq.~\eqref{gaugetf}.

Next, we will compute $\tilde \rho_K$. 
The gauge transform also applies to the coordinate time $t \rightarrow \hat t$, which consequently defines a time-like normal for the new hypersurface,  $\tilde n_a$.
The fact that we change the constant-time hypersurface in the gauge transform naturally implies
\begin{align}
    \tilde n_a\neq \hat n_a,
\end{align}
which can be shown by explicitly calculating the components of the 1--form $\tilde n_a$, an example of which we show in Section~\ref{sec:sync}.
Then, the spatial metric of the new constant-$\hat t$ hypersurface is
\begin{align}
    \tilde \gamma_{ab}\equiv \hat g_{ab}+\tilde n_a \tilde n_b, \label{gammatilde}	
\end{align}
and the extrinsic curvature in the new hypersurface is
\begin{align}
	\tilde K_{ab} \equiv \frac12\pounds_{\tilde n} \tilde \gamma_{ab}\label{defKab11tilde}.
\end{align}
The GW energy density associated with the new hypersurface, $\tilde \rho_K$, can then be calculated from $\tilde K_{ab}$ after applying the covariant TT projection \eqref{covAab} with respect to $\tilde \gamma_{ab}$.

From Eqs.~\eqref{defKab11tilde} and \eqref{kab22}, one finds $\hat K_{ab}\neq \tilde K_{ab}$ and consequently $\tilde K^{ab}_{\rm TT}\neq \hat K^{ab}_{\rm TT}$, which further implies $\hat \rho_K\neq \tilde \rho_K$.
We emphasize again that the energy density $\hat \rho_K$ is physically identified with $\rho_K$ via Eq.~\eqref{gaugetf} but \textit{does not} represent the energy density associated with the new constant-time hypersurface: it represents the energy density associated with the original hypersurface \textit{as seen from the new gauge}. 
Instead, $\tilde \rho_{K}$ is the energy density on the new constant-time hypersurface --- and therefore contains the TT component associated with $\tilde \gamma_{ab}$ and not the original spatial metric $\gamma_{ab}$.
The transformation $\rho_K\to \tilde \rho_K$ is thus \textit{not} a gauge transformation since they are physically different quantities.
In Figure~\ref{fig1}, we illustrate the relation between the different hypersurfaces involved in the transformation.

Due to the explicit dependence of $\rho_K$ on a particular spatial hypersurface, its connection to physical observables is unclear. We leave an investigation into this connection to future work. However, $\rho_K$ may still prove useful in analytic studies in cosmology via perturbation theory --- which typically need to define a spatial hypersurface. In the following section we will consider its use in both the Newtonian and synchronous gauges.

\section{Second order perturbation theory}
\label{sec:2nd-order}

In the previous section, we presented the gauge invariance of $\rho_K$ at the lowest order of $\xi$ in a general way.
Although $\rho_K$ may be computed in any gauge, it depends on the choice of a hypersurface from which to define the expansion, $K$, and thus the energy densities in Eq~\eqref{eq:ham_recast_full}. 
Therefore, identifying the hypersurface where $\rho_K$ behaves as physical GW radiation is important.
One approach is to make use of NR simulations, for which we must specify a hypersurface for the evolution of Einstein's equations. We can then directly compute $\rho_K$ from the simulation itself, i.e. without a perturbative expansion, and see if it behaves as radiation in terms of cosmological evolution.
We perform this test in a paper in preparation \citep{cmo:2021b}, while in the following we will instead consider an approach based on perturbation theory.

\medskip

We consider the metric in 3+1 form,
\begin{align}
	ds^2 = -\alpha^2dt^2 + \gamma_{ij} (\beta^idt+dx^i) (\beta^jdt+dx^j),
\end{align}
where $\alpha$ is the lapse function, $\beta^i$ is the shift vector, and $x^\mu=(t,x^i)$ are the spacetime coordinates.
The normal vector to the constant-time hypersurface is $n^\mu =\alpha^{-1} (1  , -\beta^i)$, and the components of the extrinsic curvature are
\begin{align}
	     K^{ij} = -\frac{1}{2\alpha}(\dot \gamma^{ij} 
	 -
	 \beta^k \partial_k \gamma^{ij} 
	+ \gamma^{ik} \partial_k \beta^j
		+ \gamma^{k j} \partial_k \beta^i 
),\label{Kijdef}
\end{align}
where an overdot implies a derivative with respect to coordinate time, we have defined $\partial_k \equiv \partial/\partial x^k$, and we have $K^{00}=K^{0i}=\gamma^{00}=\gamma^{0i}=0$.

In Section~\ref{sec:Newtonian}, we will first calculate $\rho_K$ in the Newtonian gauge up to fourth order in scalar perturbations. We will then consider a gauge transformation from the Newtonian gauge to the synchronous gauge in Section~\ref{sec:sync}. Consequently, we will show that the gauge-transformed energy density $\hat\rho_K$ is equivalent to $\rho_K$ at leading order, however, the energy density on the new (synchronous) constant-time hypersurfaces, $\tilde\rho_K$, is physically different.

\subsection{Newtonian gauge}\label{sec:Newtonian}

First, we consider an FLRW background with scalar perturbations and second-order induced vector and tensor perturbations in the Newtonian gauge.
Our metric ansatz is 
\begin{align}\label{newtonian}
    \alpha =e^{\phi},\quad\beta_i=C^{t}_{i},~
	\gamma_{ij} = a^2e^{2\psi}(\delta_{ij}+h^{tt}_{ij}),
\end{align}
where $a(t)$ is the background scale factor, and the superscripts ``$t$'' and ``$tt$'' imply the non-covariant transverse condition, $\partial_i C^{t}_{i}=0$, and transverse-traceless condition, $\delta_{ij}h^{tt}_{ij}=\partial_i h^{tt}_{
ij} =0$, respectively. 
The tensor perturbation $h^{tt}_{ij}$ is the second-order GWs \textit{induced} by the scalar perturbations --- composed of terms quadratic in the first order $\phi$ and $\psi$. Similarly, the vector perturbation $C^t_i$ is also of quadratic order in the scalar perturbations. 
In Eq.~\eqref{newtonian}, gauge fixing is perfect up to second order for both vector and scalar perturbations.

\medskip
Expanding Eq.~\eqref{Kijdef} up to second order for the metric \eqref{newtonian}, we find
\begin{align}
	K^{ij}	& = - \frac{1}{2} \dot \gamma^{ij}
	-\frac{1}{2a^{4}}\partial_j C^{t}_i
	-\frac{1}{2a^{4}}\partial_i C^{t}_j,\label{Kijnewtoninan} \\
	\frac{K}{3} & =H+\dot \psi.
\end{align}
Then, the trace-free part of the extrinsic curvature (the shear tensor) is evaluated as 
\begin{align}
	K^{ij}- \gamma^{ij}\frac{K}{3} = 
\frac{1}{2a^{2}}\dot h^{tt}_{ij}
	-\frac{1}{2a^{4}}\partial_j C^{t}_i
	-\frac{1}{2a^{4}}\partial_i C^{t}_j. 
\end{align}
Applying the projection in Eq.~\eqref{decom}, we find the following covariant TT decomposition of the extrinsic curvature
\begin{align}
		K_{\rm TT}^{ij} = \frac{\dot h^{tt}_{ij}}{2a^2}  , \quad V^i = -\frac{C^{t}_{i}}{2a^2},\label{Aijnewton}
\end{align}
and we have $K^{00}_{\rm TT}=K^{0i}_{\rm TT}=0$.
Hence, the leading order energy density is 
\begin{align}
	\rho_K = \frac{M^2_{\rm pl}}{8}\dot h^{tt}_{ij}\dot h^{tt}_{ij}+\cdots,\label{rhoknew}
\end{align}
which is correct to fourth order in scalar perturbations.
This is the energy density of GWs in the constant-time hypersurface in Newtonian gauge.
This expression may appear similar to the Isaacson energy density, however, this coincidence is only explicitly valid for the special case of linear tensor modes in a vacuum spacetime. For the case considered here --- namely, second-order induced tensor modes --- $\rho_K$ does not in general give the same result as the Isaacson energy density.

It has been shown that the RHS of Eq.~\eqref{rhoknew} has physically nice properties in Newtonian gauge, namely that the short-scale induced $\dot h^{tt}_{ij}$ contains only oscillating modes for various constant equations of state~\cite{Domenech:2019quo} \textit{including during matter domination}~\cite{Ali:2020sfw}.
These authors find $(\dot h^{tt}_{ij})^2 \propto a^{-4}$ in the Newtonian gauge on subhorizon scales, which further implies $\rho_{K}\propto a^{-4}$. We expect this result for propagating GWs since they should decay like radiation in an expanding universe.
Therefore, the constant-time hypersurface in Newtonian gauge could be a helpful reference to describe physical GWs.
The gauge transformation of Eq.~\eqref{rhoknew} is given by Eq.~\eqref{rhoTTgt} and the energy density is gauge invariant at fourth order in the scalar perturbations as we discussed in Section~\ref{sec:GWgauge}.
We should note that the particular form of Eq.~\eqref{rhoknew} that we find is valid only in the Newtonian gauge.

\subsection{Synchronous gauge}
\label{sec:sync}
Next, we will compute the GW energy density in the new constant-time hypersurface in the synchronous gauge, i.e., $\tilde \rho_K$.
The metric tensor in the synchronous gauge is written as
\begin{align}
	&\tilde \alpha = \hat g_{00} =-1,~	\tilde \beta_i = \hat g_{0i} = C^{S,t}_{i},
	\notag \\
	&\tilde \gamma_{ij} = \hat g_{ij} =a^2e^{2\psi^S}(\delta_{ij}+h^{S,tt}_{ij}) +2 a^2 \partial_i \partial_j E^{S},
	\label{eq:ghat}
\end{align}
where we distinguish the perturbations in the synchronous gauge from those in Newtonian gauge using the superscript $S$.
In this gauge, after the equivalent calculation as Eqs.~\eqref{Kijnewtoninan} to \eqref{Aijnewton}~(see Appendix~\ref{app:der} for details), we find
\begin{align}
		\tilde K_{\rm TT}^{ij} &= \frac{\dot{h}^{S,tt}_{ij}}{2a^2} - \frac{\dot X^{tt}_{ij}}{2a^2},\label{sync:tildeKijTT}\\
		\tilde V^i &=\frac{\partial_i \dot E^{S}}{2} -\frac{\dot X^t_i}{2}
			- \frac{\partial_k \dot E^S \partial_k \partial_i E^S}{2}
		 -\frac{C^{S,t}_{i}}{2a^2},\label{Aijnewtonsync}
\end{align}
where $X_{ij}^{tt}$ and $X_{i}^{t}$ are defined such that
\begin{align}
    &	4 \psi^{S}\partial_j\partial_i E^{S}
	+   \partial_k \partial_i E^{S}\partial_j\partial_k E^{S} \notag \\
	&= \bar X\delta_{ij} +2\partial_i\partial_j X +\partial_i X_j^t+\partial_j X_i^t + X^{tt}_{ij}.
	\label{def:X}
\end{align}
This calculation is more complicated than in Newtonian gauge because in this gauge we have a non-vanishing scalar shear, $E^S$.
From Eq.~\eqref{sync:tildeKijTT}, we find
\begin{align}
	\tilde \rho_K = \frac{M_{\rm pl}^2}{8}\left(\dot h^{S,tt}_{ij} - \dot X^{tt}_{ij} \right)^2
	,\label{rhosync}
\end{align}
from which we can see we have an additional contribution, $X_{ij}^{tt}$, to $\rho_K$ in the synchronous gauge hypersurface.

Next we wish to directly compare $\rho_K$ and $\tilde\rho_K$.  
The second-order gauge transformation of the metric tensor from the Newtonian gauge to the synchronous gauge gives the relation between the tensor modes in each gauge to be (see Appendix~\ref{appGT} for details)
\begin{align}
    h^{S,tt}_{ij} = h^{tt}_{ij} + X^{tt}_{ij}  -Y^{tt}_{ij},\label{hathij}
\end{align}
where we have introduced
\begin{align}
Y_{ij} =
	a^2 \partial_i \dot{E}^{S}   \partial_j \dot {E}^{S}.\label{def:Ymain}
	\end{align}
Combining Eqs.~\eqref{hathij} and \eqref{rhosync} we can write $\tilde\rho_K$ in terms of Newtonian gauge variables, which gives
\begin{align}
	\tilde \rho_K = \frac{M_{\rm pl}^2}{8}\left(\dot h^{tt}_{ij} - \dot Y^{tt}_{ij} \right)^2\neq \rho_{K},
	\label{rhosyncinnew}
\end{align}
i.e. the energy density in the new hypersurface (synchronous) is \textit{not equal} to the energy density on the original hypersurface (Newtonian).
The GW energy density in the synchronous gauge hypersurface contains the secondary effect of the scalar shear $E^{S}$ via $Y_{ij}$ --- which are not GWs. 
To remove these fictitious tensor modes, we must correctly identify the constant time hypersurface of the Newtonian gauge from the synchronous gauge, i.e., we must calculate $\hat \rho_K$ instead of $\tilde\rho_K$. 
Specifically, this requires using $\hat \gamma_{\mu\nu}$ instead of $\tilde \gamma_{\mu\nu}$, i.e., calculating $\hat K^{ij}_{\rm TT}$ instead of $\tilde K^{ij}_{\rm TT}$. 
Using Eq.~\eqref{gaugetf} for $n_\mu=-\alpha \delta_{0\mu}$, we find
\begin{align}
	\hat n_i = a^2 \partial_i \dot E^S.
\end{align}
Thus, $\tilde n_\mu\neq \hat n_\mu$ since $\tilde n_\mu =- \delta_{0\mu}$ and we get
\begin{align}
	\hat n_i \hat n_j - \tilde n_i\tilde n_j=& a^4 \partial_i \dot E^S \partial_j \dot E^S.
\end{align}
The two spatial metrics are thus related by
\begin{align}
    \hat \gamma_{ij} = \tilde \gamma_{ij} +  a^2 Y_{ij}.	
\end{align}
This connection implies that for the correct calculation, we should replace $h^{S,tt}_{ij} \to h^{S,tt}_{ij} +Y_{ij}^{tt} $ in Eq.~\eqref{rhosync} and thus in Eq.~\eqref{rhosyncinnew}. We thus arrive at $\hat \rho_K = \rho_{K}$ correct to fourth order in the scalar perturbations.
Thus, we confirmed Eq.~\eqref{igwgaugeinv} for this particular gauge transformation.

In the separate case of linear tensor modes, from Eq.~\eqref{def:Ymain} we will have $Y_{ij}=0$, which implies all energy densities are equivalent $\rho_{K}=\hat \rho_{K} =\tilde \rho_{K}$.
Therefore, the GW energy density in \textit{linear theory} is not only gauge independent but also hypersurface independent to leading order.

Let us briefly consider when the hypersurface dependence is negligible at second order, i.e., in which cases we find $\rho_K \approx \tilde \rho_K$.
During the radiation era, $\dot Y_{ij}$ is decreasing as we approach the sub-horizon limit such that $\rho_K=\tilde \rho_K=\hat \rho_K$~\cite{Inomata:2019yww,DeLuca:2019ufz}.
During matter domination, the formation of structures leads to natural time variation between different locations in spacetime.
The synchronous gauge unnaturally synchronizes the time and consequently the scalar shear introduces unphysical effects.
As the scalar shear grows during matter domination, so does the additional contribution to \eqref{rhosyncinnew} and we have $\tilde \rho_K \neq\hat \rho_K$. 
This argument is limited to the synchronous gauge, however, extending this to a general gauge condition should be possible.

\subsection{Analogy with gauge-invariant scalar perturbations}

We may draw an analogy between the gauge invariance of $\rho_K$ within perturbation theory and the well-known gauge invariant Bardeen potentials in cosmology \citep{Bardeen:1980}.
One usually finds the ``gauge transformation'' of metric perturbations by comparing the metric components before and after a gauge transformation.
We can then construct gauge invariant variables, such as the Bardeen potentials, by making combinations such that gauge dependent parts cancels one other.
We can also construct these variables more concicely from a geometrical perspective, as we will now demonstrate. 
In linear perturbation theory, we may define the \textit{metric perturbation tensor} as the difference between the physical metric tensor $g_{ab}$ and a fictitious background metric tensor $g_{ab(0)}$, i.e.,
\begin{align}
    \delta g_{ab} \equiv g_{ab} - g_{ab(0)}\label{mpt}
\end{align}
where the parentheses indicate the order of each tensor.
Eq.~\eqref{mpt} is different from ``metric perturbations'' we defined in Eqs.~\eqref{newtonian} and \eqref{eq:ghat}.
Once we fix a reference gauge, we always identify the equivalent background metric $g_{ab(0)}$, which is obtained by transforming $g_{ab(0)}$ with Eq.~\eqref{gaugetf}.
As a result, $\delta g_{ab}$ is also considered as a spacetime tensor.
To linear order, the gauge transformation of $\delta g_{ab}$ is written as 
\begin{align}
	\delta g_{ab} \to \delta g_{ab} + \pounds_\xi \delta g_{ab}.
\end{align} 
Thus, the \textit{metric perturbation tensor} is a linear gauge invariant, which is a formal construction of a gauge-invariant quantity.
In Newtonian gauge, the first order $\delta g_{ab(1)}$ is constructed from the Bardeen potentials $\Phi_B$ and $\Psi_B$. Specifically, in component representation we have 
\begin{align}
	\delta g_{00(1)}& = -2\Phi_B,\\
	\delta g_{ij(1)}&= 2a^2 \Psi_B\delta_{ij},\label{BD}
\end{align}
and $\delta g_{0i(1)}=0$. 
\textit{Only} in this gauge we find that $\Phi_B$ and $\Psi_B$ coincide with the gravitational potential and curvature perturbations, respectively. However, it is well known that we can always compute the Bardeen potentials as a combination of the perturbations in a generic gauge.
Thus, the metric perturbation tensor is gauge invariant to the lowest order as we simultaneously transform both $g_{ab}$ and $g_{ab(0)}$.
As shown in Section~\ref{sec:2nd-order}, this is analogous to our formalism within perturbation theory. Namely, we used the constant-time hypersurface in the Newtonian gauge as a reference, and maintaining the same reference for the GW energy density before and after the gauge transformation we found that $\rho_K$ is gauge invariant.

\section{Conclusions}

The effective GW energy momentum tensor in the literature is gauge dependent for the second-order scalar-induced GWs.
This implies that 
it is not consistently defined from a spacetime tensor.
This is a crucial problem for the physical interpretation of these GWs since any physical quantities must be written as tensors in the physical spacetime. 
In this paper, we revisited the definition of GWs outside of perturbation theory and showed that the covariant TT part of the extrinsic curvature may represent the kinetic energy of GWs associated with a particular hypersurface.
The definition is based only on spacetime tensors. We showed that we can correctly interpret the gauge transformation of the GW energy density by identifying the original hypersurface on which the GW energy density of interest was defined.
Our work is consistent with the traditional Isaacson formalism for linear GWs and may be straightforwardly used in analyses of second-order induced GWs.
This new energy density has gauge invariance at leading order by construction, and we have shown an example of this by calculating its gauge transformation from the Newtonian to synchronous gauge.
Our approach is independent of any form of the stress-energy tensor and therefore is valid for eras of radiation, matter, or dark energy dominance (or any combination of these).

We propose Eq.~\eqref{eq:rhoK} as a non-perturbative characterization of the kinetic energy density of GWs in the expanding Universe, with the gauge freedom wholly removed.

While $\rho_{K}$ is independent of any particular gauge choice (i.e., a map between the physical spacetime and a fictitious background), it is explicitly dependent on a particular choice of spatial hypersurface. 
The issue of finding hypersurfaces which best represent the physical GWs that we observe remains to be solved \cite[though see][]{Domenech:2020xin}. Additionally, the relation of $\rho_{K}$ (as defined on a hypersurface) to the {observable} signature of GWs remains unclear, and we leave this to future work.
However, we have shown that
the Newtonian constant-time hypersurface could be useful to study physical (purely oscillating) GWs at second order. We have not explored the extension of $\rho_K$ to a fully nonlinear framework in this paper. We investigate this extension, making use of NR simulations of nonlinear cosmological structure formation, in a paper in preparation~\cite{cmo:2021b}.

\acknowledgments
 We thank the anonymous referees whose comments improved the quality and clarity of the manuscript.
The authors would also like to thank Katy Clough, William Barker, Misao Sasaki, and Keisuke Inomata for useful comments on a draft of this letter, and Chao Chen, Kouji Nakamura and Guillem Dom\`enech for helpful discussions. HJM appreciates the support received from the Herchel Smith Postdoctoral Fellowship Fund. The Flatiron Institute is supported by the Simons Foundation.

\appendix

\section{Index convention}
\label{index:spp}
In this paper, we distinguish between the spatial component of tensors (defined with a coordinate basis) and variables labeled by the spatial index (which are not tensors or components of tensors).

As an example of the former, $K_{\mu\nu}$ represents the components of the spacetime tensor $K_{ab}$ with a coordinate basis (whereas $K_{ab}$ is defined without reference to any coordinate basis).
$K_{ij}$ then represents the spatial components of $K_{\mu\nu}$.
The spatial upper indices in $K^{ij}$ imply the spatial part of $K^{\mu\nu}$, which is the component of the corresponding tensor in the cotangent space, $K^{ab}$.
The extrinsic curvature is automatically projected onto the hypersurface, so we have
\begin{align}
	K^{ij} =g^{i\mu}g^{j\nu} K_{\mu\nu}=\gamma^{i\mu}\gamma^{j\nu} K_{\mu\nu}, 
\end{align}
where the last equality is true only for the tensors projected onto the hypersurface.
The index levels are meaningful only for tensors to distinguish the tangent space and the cotangent space.

On the other hand, quantities such as the vector and tensor perturbations $C_i^t$ and $h^{tt}_{ij}$, respectively, and the partial derivative $\partial_i$ are \textit{variables labeled by the spatial indices}, which are not components of tensors themselves.
In this paper, we do not make use of the Kronecker delta $\delta_{ij}$ to raise and lower the indices of quantities which are not tensors. Instead, we clarify that we have contraction/summation over repeated indices \textit{regardless of position}. We choose to have raised indices only for quantities which are tensors.

\begin{widetext}

\section{Covariant TT decomposition in Synchronous gauge}
\label{app:der}
In this section we provide a derivation for the TT decomposition of the extrinsic curvature associated with the synchronous gauge hypersurfaces, namely Eqs.~\eqref{sync:tildeKijTT} and \eqref{Aijnewtonsync}.
We first compute the extrinsic curvature using Eq.~\eqref{Kijdef} for the spatial metric in synchronous gauge \eqref{eq:ghat} and derive the traceless part of the extrinsic curvature (shear tensor).
Then, we define $\tilde V^i$ in such a way that the transverse component of the shear tensor is all subtracted.
In this section, we always truncate the perturbative expansion at second order in scalar perturbations.
The result for $K^{ij}_{\rm TT}$ in Newtonian gauge shown in Eq.~\eqref{Aijnewton} can be found by setting $E^S=0$ in this derivation.

\medskip
First of all, from Eqs.~\eqref{Kijdef} and \eqref{eq:ghat}, we obtain	
\begin{align}
	\tilde K^{ij} 
	& = - \frac{1}{2} \dot {\tilde \gamma}^{ij}
	-\frac{1}{2a^{4}}\partial_j C^{S,t}_i
	-\frac{1}{2a^{4}}\partial_i C^{S,t}_j,\label{eq:b1}
\end{align}
where the tilde implies that the induced metric and associated extrinsic curvature are defined on the constant-time hypersurface in the synchronous gauge.
The inverse of the induced metric is
\begin{align}
	\tilde \gamma^{ij} = \frac{1}{a^2}e^{-2\psi^S}(\delta_{ij}-h_{ij}^{S,tt}) -\frac{2}{a^2}\partial_i \partial_j  E^S + \frac{8}{a^2} \psi^S \partial_i\partial_j  E^S + \frac{4}{a^2}\partial_i \partial_k E^S\partial_k \partial_j E^S,
	\label{eq:b2}
\end{align}
which satisfies $\tilde \gamma^{il}\tilde \gamma_{lj} = \delta_{ij}$.
Substituting \eqref{eq:b2} into \eqref{eq:b1}, and taking the trace of the extrinsic curvature, we find
\begin{align}
	\frac{\tilde K}{3}  =H+\dot \psi^S 
	+\frac{1}{3}\partial^2  \dot E^S
	-\frac{2}{3} (  \psi^S \partial^2  E^S \dot )
	-\frac{1}{3}(\partial_l \partial_k E^S\partial_l \partial_k E^S\dot ).
	\label{eq:b3}
	\end{align}
From Eqs.~\eqref{eq:b1} to~\eqref{eq:b3}, we arrive at the following shear tensor:
\begin{align}
	\tilde K^{ij}- \tilde \gamma^{ij}\frac{\tilde K}{3} &= 
	\frac{1}{a^2} \left [ -\frac{1}{3}\partial^2  \dot E^S 	
	+\frac{2}{3}   (\psi^S \partial^2E^S\dot)
	+\frac{2}{3}   \psi^S    \partial^2  \dot E^S 
		+\frac{1}{3}(\partial_l \partial_k E^S\partial_l \partial_k E^S\dot )\right]\delta_{ij}
		\notag \\
		&+\frac{1}{2a^{2}}\dot h^{S,tt}_{ij}
	+\frac{1}{a^2} \partial_i \partial_j   \dot E^S	
	-\frac{2}{a^2} (\psi^S \partial_i \partial_j  E^S \dot)
	-\frac{2}{a^2}   \psi^S   \partial_i \partial_j  \dot E^S \notag \\
	&-\frac{2}{a^2}(\partial_i \partial_k  E^S\partial_k \partial_j E^S\dot )
	+\frac{2}{3a^2}\partial_i\partial_j E^S\partial^2  \dot E^S
	-\frac{1}{2a^{4}}\partial_j C^{S,t}_i
	-\frac{1}{2a^{4}}\partial_i C^{S,t}_j. 
	\label{b:shear:eq}
\end{align}

\medskip
Next, we compute $(L\tilde V)^i$.
We will put forward an ansatz of $\tilde V^ i$ and determine the form to subtract all transverse part from the shear tensor \eqref{b:shear:eq}.
The covariant derivative of $\tilde V^i$ is defined as
\begin{align}
	\tilde D^j \tilde V^{ i} &=\tilde \gamma^{jm} \partial_m \tilde V^i + \tilde \gamma^{jm}\tilde \Gamma^{i}_{mk}\tilde V^k,
\end{align}
where $\tilde \Gamma^i_{mk}$ and $\tilde D_i$ are the Christoffel symbol and the covariant derivative with respect to $\tilde \gamma_{ij}$, respectively.
Then we find
\begin{align}
	\tilde D^j \tilde V^{ i}+ \tilde D^i \tilde V^{ j} 
	&=\tilde  \gamma^{jm} \partial_m \tilde V^i  + \tilde \gamma^{im} \partial_m \tilde V^j 
	-\tilde V^k  \partial_k \tilde \gamma^{ij},
	\label{DVDV}
\end{align}
Substituting Eq.~\eqref{eq:b2} into Eq.~\eqref{DVDV}, we find
\begin{align}
		\tilde D^j \tilde V^ { i}+ \tilde D^i \tilde V^ { j} 
		& =\frac{1}{a^2}[
	 \partial_j \tilde V^i +  \partial_i \tilde V^j -2   \psi^S\partial_j \tilde V^i -2   \psi^S\partial_i \tilde V^j 
	 +  2\tilde V^k  \partial_k   \psi^S \delta_{ij}
	 \notag 
	 \\
	 &
	 -2\partial_k \tilde V^i \partial_j \partial_k  E^S    
	 -2\partial_k \tilde V^j \partial_i \partial_k  E^S 
	 +2\tilde V^k  \partial_k \partial_i \partial_j  E^S].
	 \label{eq:b1:b2}
\end{align}
Using \eqref{eq:b1:b2} in the definition of $(LV)^{ab}$ in Eq.~\eqref{LL}, we obtain
\begin{align}
(L\tilde V)^{ij}=&\frac{1}{a^2}[
	 \partial_j \tilde V^i +  \partial_i \tilde V^j -2   \psi^S\partial_j \tilde V^i -2   \psi^S\partial_i \tilde V^j 
	 -2\partial_k \tilde V^i \partial_j \partial_k  E^S   
	 -2\partial_k \tilde V^j \partial_i \partial_k  E^S  
	 \notag \\
	 &
	 	 +2\tilde V^k  \partial_k \partial_i \partial_j  E^S
	 + \frac{4}{3} \partial_k \tilde V^k \partial_i \partial_j  E^S
		-\frac{2}{3}\delta_{ij} (\partial_k \tilde V^k -	2  \partial_k \tilde V^k \psi^S + \tilde V^k\partial_k \partial^2 E^S)].
		\label{LYi}
\end{align}
Now, we put forward the following ansatz for $\tilde V^i$:
\begin{align}
	\tilde V^i = \frac{1}{2} \partial_i \dot E^S - \frac{1}{2}\partial_k \dot E^S \partial_k \partial_i E^S - \frac{1}{2}\dot X^t_i- \frac{1}{2a^2}C^{S,t}_i,\label{ansY}
\end{align}
where we will fix $\dot X_i^t$ later.
Substituting Eq.~\eqref{ansY} into Eq.~\eqref{LYi}, we get
\begin{align}
(L\tilde V)^{ij}
=	&\frac{1}{a^2}[
	 \partial_j \partial_i \dot E^S  -2   \psi^S \partial_j \partial_i \dot E^S  
	 -(\partial_k \partial_i  E^S \partial_j \partial_k  E^S\dot )    
	 +\partial_k \dot E^S  \partial_k \partial_i \partial_j  E^S
	 + \frac{2}{3} \partial_k \partial_k \dot E^S \partial_i \partial_j  E^S 
	 \notag \\
	 &
		-\frac{1}{3}\delta_{ij} (\partial_k \partial_k \dot E^S -	2  \partial_k \partial_k \dot E^S \psi^S + \partial_k \dot E^S\partial_k \partial^2 E^S
		-\partial_k (\partial_l \dot E^S \partial_l \partial_k E^S)
		)
		-\frac{1}{2}(\partial_k \partial_j E^S \partial_k \partial_i E^S\dot)
		-\partial_k  \dot E^S \partial_k \partial_i \partial_j  E^S]
		\notag \\
		&
		- \frac{1}{2a^2}\partial_i \dot X^t_j- \frac{1}{2a^2}\partial_j \dot X^t_i
		- \frac{1}{2a^4}\partial_i C^{S,t}_j- \frac{1}{2a^4}\partial_j C^{S,t}_i.
		\label{LYi2}
\end{align}
Finally, combing Eqs.~\eqref{b:shear:eq} and \eqref{LYi2}, we get
\begin{align}
	\tilde K^{ij}- \tilde \gamma^{ij}\frac{\tilde K}{3}-(L\tilde V)^{ij} &= \frac{1}{2a^{2}}\dot h^{S, tt}_{ij}
	+
	\frac{1}{2a^2}\partial_i \dot X^t_j+\frac{1}{2a^2} \partial_j \dot X^t_i
	\notag \\
	&+\frac{1}{a^2}
	\Bigg(2   \psi^S \partial^2E^S
		+\frac{1}{2}\partial_l \partial_k E^S\partial_l \partial_k E^S
	\dot {\Bigg)}
	\frac{\delta_{ij}}{3}- \frac{1}{a^2}\Bigg(2   \psi^S \partial_i\partial_j E^S
		+\frac{1}{2}\partial_i \partial_k E^S\partial_j \partial_k E^S
	\dot {\Bigg)}.
\end{align}
Now, we introduce $X_{ij}^{tt}$ and $X^t_i$ via Eq.~\eqref{def:X} in the main text.
Then, from Eq.~\eqref{ansY} we arrive at the result for $\tilde V^i$ given in Eq.~\eqref{Aijnewtonsync}, and we thus obtain Eq.~\eqref{sync:tildeKijTT} via
\begin{align}
	\tilde K^{ij}_{\rm TT} = \tilde K^{ij}- \tilde \gamma^{ij}\frac{\tilde K}{3}-(L\tilde V)^{ij} 
	&= 	\frac{\dot h^{tt}_{ij}}{2a^{2}} - \frac{\dot X^{tt}_{ij}}{2a^2}.
\end{align}

\section{Gauge transformation of the spacetime metric}\label{appGT}
The gauge transformation~\eqref{gaugetf} expands to second order as~\cite{Sonego:1997np}
\begin{align}
	\hat Q_0  =\left(1 +  \pounds_{\xi_{(1)}} + \pounds_{\xi_{(2)}} + \frac{1}{2}\pounds_{\xi_{(1)}}^2+\cdots \right)Q_0.\label{secondgt}
\end{align}
In this appendix, we apply this transformation Eq.~\eqref{secondgt} for the metric tensor $g_{\mu\nu}$ from the Newtonian gauge to the synchronous gauge.
We specify the order in scalar perturbations with a subscript in parentheses.

\subsection{First order}
At first order, the gauge transformation is
\begin{align}
	\delta_{(1)} g_{00}	& = -2 \dot \xi^0_{(1)}
	\\
	\delta_{(1)} g_{0i}	& = -\partial_i\xi^0_{(1)}+ a^2 \dot \xi^i_{(1)}
	\\
	\delta_{(1)} g_{ij}	& = 2a^2 H \xi^0_{(1)}\delta_{ij} + a^2 \partial_j\xi^i_{(1)}+ a^2\partial_i\xi^j_{(1)}.
\end{align}
To realize the transformation from Newtonian gauge to the synchronous gauge, the first order $\xi$ must satisfy
\begin{subequations}\label{eqs:1stxi_constraints}
\begin{align}
	0	& = -\phi_{(1)}- \dot \xi^0_{(1)},
	\\
	0	& = -\partial_i\xi^0_{(1)}+ a^2 \dot \xi^i_{(1)}\label{ge0i},
	\\
	\psi^S_{(1)} &= \psi_{(1)} +H \xi^0_{(1)},\\
	2a^2 \partial_i\partial_j E^S_{(1)}	& = a^2 \partial_j\xi^i_{(1)}+ a^2\partial_i\xi^j_{(1)}.\label{geij}
\end{align}
\end{subequations}
From Eqs.~\eqref{ge0i} and \eqref{geij}, we get
\begin{align}
		\xi^i_{(1)} &= \partial_i E^S_{(1)},\\
		 \xi^0_{(1)} &= a^2 \dot {E}^S_{(1)},
\end{align}
which implies that $E^S$ satisfies $(a^2\dot{E}^S_{(1)}\dot) = -\phi_{(1)}$.

\subsection{Second order}
The second-order gauge transformation from Newtonian gauge to the synchronous gauge is
\begin{align}
	\delta_{(2)} g_{00}	& = -2 \dot \xi^0_{(2)}- \xi^\rho_{(1)} \partial_\rho  \dot \xi^0_{(1)} -2 \dot \xi^0_{(1)}\dot\xi^0_{(1)} -2 \xi^\rho_{(1)} \partial_\rho \phi_{(1)} -4 \phi_{(1)}\dot \xi^0_{(1)}
	\\
	\delta_{(2)} g_{0i}	& = -\partial_i\xi^0_{(2)}+ a^2 \dot \xi^i_{(2)}
	- \frac{1}{2}\xi^\rho_{(1)} \partial_\rho ( \partial_i\xi^0_{(1)}- a^2 \dot \xi^i_{(1)} ) 
	- (\phi_{(1)}+\dot \xi^0_{(1)})\partial_i\xi^0_{(1)}+ a^2 (\psi_{(1)}+H \xi^0_{(1)})\dot\xi^i_{(1)} \notag \\
	&+ \frac{1}{2} a^2 \partial_k \xi^i_{(1)}\dot\xi^k_{(1)}+ \frac{1}{2} a^2\partial_i\xi^k_{(1)}\dot \xi^k_{(1)}\notag \\
	&-\phi_{(1)}\partial_i\xi^0_{(1)}+ a^2 \psi_{(1)}\dot \xi^i_{(1)} 
	\\
	\delta_{(2)} g_{ij}	& =\left[ 2a^2 H \xi^0_{(2)}  +2 \xi^\rho_{(1)} \partial_\rho (a^2\psi_{(1)})
	+
	\xi^\rho_{(1)} \partial_\rho  (a^2 H \xi^0_{(1)}) \right] \delta_{ij}
	+ a^2 \partial_j\xi^i_{(2)}+ a^2\partial_i\xi^j_{(2)}  
\notag 	\\
	&+ 2a^2 (H \xi^0_{(1)} + \psi_{(1)})\partial_j\xi^i_{(1)}+ 2 a^2 (H \xi^0_{(1)} + \psi_{(1)}) \partial_i\xi^j_{(1)}
	\notag \\
	&
	+ \frac{1}{2} a^2 \partial_k\xi^i_{(1)}\partial_j\xi^k_{(1)}+  \frac{1}{2}a^2\partial_k\xi^j_{(1)}\partial_i\xi^k_{(1)}
	+ a^2\partial_i\xi^k_{(1)}\partial_j\xi^k_{(1)}\notag \\
	&+
	\frac{a^2}{2}\xi^0_{(1)}   \partial_j \dot \xi^i_{(1)}+
	\frac{a^2}{2}\xi^0_{(1)}  \partial_i \dot \xi^j_{(1)}
	+
	\frac{a^2}{2}\xi^k_{(1)} \partial_k  \partial_j\xi^i_{(1)}+
	\frac{a^2}{2}\xi^k_{(1)} \partial_k \partial_i\xi^j_{(1)}.
\end{align}	
Using the constraints at first order in \eqref{eqs:1stxi_constraints}, the second-order gauge transformation is simplified to
\begin{align}
	\delta_{(2)} g_{0i}	& =  -\partial_i\xi^0_{(2)}+ a^2 \dot \xi^i_{(2)}
	 +  a^2 \partial_k \partial_i E^S_{(1)}\partial_k \dot {E}^S_{(1)}+ a^2(2\psi^S_{(1)}+a^2 H\dot {E}^S+a^2\ddot {E}^S)\partial_i \dot {E}^S_{(1)},\\
		\delta_{(2)} g_{ij}	
	& =\left[ 2a^2 H \xi^0_{(2)}  +2 \xi^\rho_{(1)} \partial_\rho \psi_{(1)}
	+
	\xi^\rho_{(1)} \partial_\rho  (a^2 H \xi^0_{(1)}) \right] \delta_{ij}
	+ a^2 \partial_j\xi^i_{(2)}+ a^2\partial_i\xi^j_{(2)} 
\notag 	\\
	& + 4a^2 \psi^S_{(1)}\partial_j\partial_i E^S_{(1)}
	+ 2 a^2 \partial_k \partial_i E^S_{(1)}\partial_j\partial_k E^S_{(1)}+
	a^4\dot{E}^S_{(1)}  \partial_i \partial_j \dot {E}^S_{(1)}
	+
	a^2\partial_k E^S_{(1)} \partial_k \partial_i \partial_j E^S_{(1)}
	.
\end{align}
Next, we consider the following decomposition:
\begin{align}
	\partial_k \partial_i E^S_{(1)}\partial_k \dot {E}^S_{(1)}+ (2\psi^S_{(1)}+a^2H\dot {E}^S+a^2\ddot {E}^S)\partial_i \dot {E}^S_{(1)} = \partial_i Z + Z^t_i,
\end{align}
and
\begin{align}
	&4 \psi^S_{(1)}\partial_j\partial_i E^S_{(1)}
	+ 2  \partial_k \partial_i E^S_{(1)}\partial_j\partial_k E^S_{(1)}+
	a^2 \dot{E}^S_{(1)}  \partial_i \partial_j \dot {E}^S_{(1)}
	+
	\partial_k E^S_{(1)} \partial_k \partial_i \partial_j E^S_{(1)} \notag \\
	=& \bar W\delta_{ij} + 2\partial_i\partial_j W  + \partial_i W^t_j + \partial_j W^t_i + W^{tt}_{ij}.
\end{align}
To realize the synchronous gauge up to second order, the second-order tangents, $\xi$, must satisfy
\begin{align}
	\xi^i_{(2)} &= \partial_i E^S_{(2)}- \partial_i W  - W^t_{i},
	\\ \xi^0_{(2)} &= a^2  \dot {E}^S_{(2)}  - a^2 \dot W+  a^2 Z.
\end{align}
The gauge transformation for the secondary vector and tensor perturbations are given as 
\begin{align}
	C^{S,t}_{(2)i} & = C^t_{(2)i} -aW^t_i + aZ^t_i ,\\
	h^{S,tt}_{(2)ij} & = h^{tt}_{(2)ij} + W^{tt}_{ij},
\end{align}
and integrating by parts we also find
\begin{align}
	\left[4 \psi^S_{(1)}\partial_j\partial_i E^S_{(1)}
	+   \partial_k \partial_i E^S_{(1)}\partial_j\partial_k E^S_{(1)}-
	a^2 \partial_i \dot{E}^S_{(1)}   \partial_j \dot {E}^S_{(1)}
	\right]_{tt}= W^{tt}_{ij},
\end{align}
where, in the main text, we have used $W^{tt}_{ij}= X^{tt}_{ij} - Y^{tt}_{ij}$.

\end{widetext}

\bibliography{bib}{}
\bibliographystyle{unsrturl}

\end{document}